\documentclass{ws-mpla}
\usepackage[super,compress]{cite}
\usepackage{graphicx}
\begin{document}
\markboth{V. V. Khruschov}
{Interpretation of the XENON1T excess in the model with decaying sterile neutrinos}

\catchline{}{}{}{}{}
%
%%%%%%%%%%%%%%%%%%%%%%%%%%%%%%%%%%%%%%%%%%%%%%%%%%%%%%%%%%%%%%%%%%%%

\title{Interpretation of the XENON1T excess in the model with decaying sterile neutrinos}

\author{V. V. Khruschov$^{\ast
,\$}$ 
}

\address{$^{\ast}$National Research Centre Kurchatov Institute, Academician
Kurchatov Place 1, Moscow, 123182 Russia
\\
$^{\$}$khruschov\_vv@nrcki.ru
}

\maketitle

\begin{history}
\received{Day Month Year}
\revised{Day Month Year}
\end{history}

\begin{abstract}
The phenomenological model with three active and three sterile neutrinos is used for
interpretation of the observed XENON1T excess of electronic recoil events in the 1 -- 7 keV energy region. Assuming two sterile neutrinos with appropriate mass values decay while the third sterile neutrino is stable it is possible to explain the observed energy spectrum of electronic recoil events. Moreover using this approach three peaks in the 1 -- 7 keV energy region are predicted. Dark bosons have to mix to only a small extent with photons which can be emitted in this region. The possible existence of the three light sterile neutrinos may have  perceptible influence on some phenomena in neutrino physics, astrophysics and cosmology.

\end{abstract}

\ccode{PACS numbers: 12.10.Kt, 12.90.+b, 14.60.Pq, 14.60.St}

\section{Introduction}
\label{Section_Introduction}
Recently the XENON1T experiment data concerning electronic recoil events in the energy region between 1 and 210 keV have been reported \cite{apr}.  The excess of electronic recoil events 
in the energy region between 1 and 7 keV has been observed using the data. At the moment there is an increasing list of papers related to explanation for the excess (e.g. \cite{bell,choi,lind,ge,amin,ch,ar,sen,moh,far}). In the present  paper we suggest 
an interpretation of the observed effect in the framework of the neutrino model with three active and three sterile neutrino \cite{khfo} with decaying two sterile neutrinos.

It is known that oscillations of solar, atmospheric, reactor and accelerator active neutrinos  can be attributed to mixing of three mass states of neutrinos that is effected by way of the Pontecorvo--Maki--Nakagawa--Sakata matrix $U_{\rm PMNS}\equiv U = V\!P$, so that 
$\psi_a^L=\sum_iU_{ai}\psi_i^L$, where $\psi_{a,i}^L$ are left chiral fields with flavor $a$ or mass $m_i$, $a=\{e,\mu,\tau\}$ and $i=\{1,2,3\}$.  The matrix
$V$ is expressed in the standard parametrization \cite{PDG} for three active neutrinos 
via the mixing
angles $\theta_{ij}$ and the CP-phase, namely, the phase
$\delta\equiv\delta_{\rm CP}$ associated with CP violation in the lepton
sector for Dirac or Majorana neutrinos, and
$P={\rm diag}\{1,e^{i\alpha},e^{i\beta}\}$, where
$\alpha\equiv\alpha_{\rm CP}$ and $\beta\equiv\beta_{\rm CP}$ are phases
associated with CP violation only for Majorana neutrinos. 

With the help of high-precision experimental data, the values of the mixing
angles $\theta_{ij}$ and the differences of the neutrino masses in square $\Delta m_{21}^2$
and $\Delta m_{31}^2$ were found \cite{PDG,salas} (where
$\Delta m_{ij}^2=m_i^2-m_j^2$), but only absolute value of $\Delta m_{31}^2$ is known, therefore, the absolute values of the neutrino masses can be ordered by two ways, namely, as
$m_1<m_2<m_3$ or $m_3<m_1<m_2$ which are called as normal neutrino mass ordering 
(NO) and as inverse neutrino mass ordering (IO), respectively. Including nonzero neutrino masses results in the Modified Standard Model (MSM) instead of the Standard Model (SM). If we take into account the data of the T2K experiment \cite{Kabe} and the limitations on the sum of the neutrino masses from cosmological observations \cite{Wang}, then the NO-case of the neutrino mass spectrum turns out to be preferable (see also \cite{salas}). However the estimation of the value of CP-phase $\delta_{\rm CP}$  \cite{salas} and the possibility of realization of the IO-case \cite{kelly} has been obtained. Nevertheless we restrict ourselves in what follows to the NO-case only, assuming $\delta_{\rm CP}=1.2\pi$.

At the same time, there are indications to anomalies of neutrino fluxes for
some processes that can not be explained with using oscillation parameters
only for three active neutrinos. These anomalies include the LSND (or accelerator)
anomaly \cite{Atha1996,Agu2001,Agu2013,Agu2018}, the reactor antineutrino anomaly
\cite{Mu2011,Me2011,Hu2011,Ko,ale18,ser18} and the gallium (or calibration)
\cite{Abdu2009,Kae2010,Giunti2013} anomaly. The anomalies manifest
themselves at short distances (more precisely, at distances $L$ such that the
numerical value of the parameter $\Delta m^2 L/E$, where $E$ is the neutrino
energy, is of the order of unity). In the LSND anomaly, an excess of the
electron antineutrinos in beams of muon antineutrinos in comparison with the
expected value according to the MSM is observed. Similar results were
observed in the MiniBooNE experiments for electron neutrinos and antineutrinos
\cite{Agu2013,Agu2018}. Deficit of reactor electron antineutrinos at short
distances is called as the reactor antineutrino anomaly, while the deficit of electron
neutrinos from a radioactive source observed at calibration of detectors for
the SAGE and GALLEX experiments is commonly called as the gallium or calibration anomaly. In
other words, data on the neutrino anomalies refer to both the appearance of the electron
neutrinos or antineutrinos excess in beams of muon neutrinos or antineutrinos,
respectively, and to the deficit of electron neutrinos or antineutrinos.
These three types of the shot-baseline (SBL) neutrino anomalies, for which
there are indications at present, are attributed to the presence of one or
two new neutrinos that do not interact directly with the gauge bosons of the
MSM, that is sterile neutrinos. The characteristic mass scale of sterile
neutrino used for explanation of the SBL anomalies is about $1$~eV.

In principle, the number of additional neutrinos can be arbitrary (see, for
example, Refs.~\citen{Bilenky1977,Abazajian2012,Bilenky}).
Phenomenological models with sterile neutrinos are usually denoted as (3+$N$) models, or,
in detail, as ($k$+3+$n$+$m$) models, where $k$ is a number of new neutrinos
with masses less than masses of active neutrinos, and $n$ and $m$ are 
numbers of new neutrinos with masses higher and considerably higher,
respectively, than masses of the active neutrinos.

 In Section~\ref{Section_OscillationModel}, the main
concepts of the (3+3) model (to be exact, the (3+1+2) model) are given, which based on the results reported in Ref.~\citen{khfo}.
In Section~\ref{xenon_ex}, we present a short description of
data relevant to the electronic recoil events excess in the XENON1T experiment and their interpretation in the context of the (3+1+2) model.
  In the final Section~\ref{Section_Conclusion} it is noticed
that the results of the present paper can help to explain the available XENON1T 
experimental data,  as well as to interpret both  data of SBL experiments on the search of sterile neutrinos and some astrophysical and cosmological data.

\section{Basic propositions of the phenomenological (3+1+2) model}
\label{Section_OscillationModel}

The (0+3+$N$) or (0+3+$m$+$n$) phenomenological neutrino models can be used to
describe the SBL anomalies, as well as some astrophysical data, where $N=m+n$
is the number of additional neutrinos. It is desirable that the number of new
neutrinos would be minimal, so the most common are the (3+1) and (3+2) models \cite{Kopp2013} ((3+1) is used instead of (0+3+1) for short).
However, if we apply the principle of extended symmetry of weak interactions,
then, for example, for the left-right symmetry it is necessary to
consider (3+3) models \cite{Conrad2013,Zysina2014,KhruFom2016}.
So, below we use the (3+1+2) model to account for effects of
light and heavy sterile neutrinos. This model includes three active neutrinos $\nu_a$
($a=e,\mu,\tau$) and three new neutrinos: a sterile neutrino $\nu_s$, a hidden
neutrino $\nu_h$ and a dark neutrino $\nu_d$. Thus six neutrino flavour states
and six neutrino mass states are present in the (3+1+2) model\cite{khfo}.
Hence below we consider the
$6\!\times\!6$ mixing matrix, which can be called as the generalized mixing
matrix $U_{\rm mix}$, or the generalized Pontecorvo--Maki--Nakagawa--Sakata
matrix $U_{\rm GPMNS}\equiv U_{\rm mix}$.
% \cite{khfo,KhruFom2016,KhruFom2017}. 
This matrix can be represented as the matrix product $V\!P$, where $P$ is a
diagonal matrix with Majorana CP-phases $\phi_i$, $i=1,\dots,5$, namely,
$P={\rm diag}\{1,e^{i\phi_1},\dots,e^{i\phi_5}\}$. We deal  only with
a particular type of matrix $V$. Keeping continuity of the notations, we
denote Dirac CP-phases as $\delta_i$ and $\kappa_j$, and 
 mixing angles as $\theta_i$ and $\eta_j$, with
$\delta_1\equiv\delta_{\rm CP}$, $\theta_1\equiv\theta_{12}$,
$\theta_2\equiv\theta_{23}$ and $\theta_3\equiv\theta_{13}$.

For the compactness of the formulas, we introduce the symbols $h_s$ and
$h_{i'}$ for left flavor fields and left mass fields,
respectively. As $s$ we will use a set of indices that allocate $\nu_s$,
$\nu_h$ and $\nu_d$ fields among $h_s$, and as $i'$ we will use a set of
indices $4$, $5$ and $6$. The common $6\!\times\!6$ mixing matrix
$U_{\rm mix}$ can then be expressed through $3\!\times\!3$ matrices $R$, $T$,
$V$ and $W$ as follows
\begin{equation}
\left(\begin{array}{c}\nu_a\\ h_s \end{array}\right)=
U_{\rm mix}\left(\begin{array}{c}\nu_i\\ h_{i'}\end{array}\right)\equiv
\left(\begin{array}{cc}R&T\\ V&W\end{array}\right)
\left(\begin{array}{c}\nu_i\\ h_{i'}\end{array}\right).
\label{eq_Umix}
\end{equation}
We represent the matrix $R$ in the form of $R=\varkappa U_{\rm PMNS}$, where
$\varkappa=1-\epsilon$ and $\epsilon$ is a small value, while the matrix $T$
in equation~(\ref{eq_Umix}) should also be small as compared with the known unitary
$3\!\times\!3$ mixing matrix of active neutrinos  $U_{\rm PMNS}$
($U_{\rm PMNS}U_{\rm PMNS}^+=I$). Thus, when choosing the appropriate
normalization, the active neutrinos mix, as it should be in the MSM,
according to Pontecorvo--Maki--Nakagawa--Sakata matrix $U_{\rm PMNS}$. 
Below we  use the notation $U_{\rm PMNS}\equiv U$.

On the current stage of the study, it is quite reasonable to restrict our consideration
to the minimal number of mixing matrix parameters that is able to explain the
available (still rather dispersive) experimental data attributed to the
SBL anomalies. The transition to full
matrix with all parameters can be done in the future, when quite reliable
experimental results will be obtained. So,  we
will consider only some particular cases, but not the most common form
for the matrix $U_{\rm mix}$.
Bearing in mind that, in accordance with data available due to astrophysical
and laboratory measurements, the mixing between active and new neutrinos is
small, we choose the matrix $T$ as $T=\sqrt{1-\varkappa^2}\,a$, where $a$ is
an arbitrary unitary $3\!\times\!3$ matrix, that is, $aa^+=I$. The matrix
$U_{\rm mix}$ can now be written in the form of
\begin{equation}
U_{\rm mix}=\left(\begin{array}{cc}R&T\\ V&W\end{array}\right)\equiv
\left(\begin{array}{cc}\varkappa U&\sqrt{1-\varkappa^2}\,a\\
\sqrt{1-\varkappa^2}\,bU&\varkappa c \end{array}\right),
\label{eq_Utilde}
\end{equation}
where $b$ is also an arbitrary unitary $3\!\times\!3$ matrix ($bb^+=I$), and
$c=-ba$. With these conditions, the matrix $U_{\rm mix}$ will be unitary
($U_{\rm mix}U_{\rm mix}^+=I$). In particular, we will use the following
matrices $a$ and $b$:
\begin{equation}
a=\left(\begin{array}{lcr}\,\,\,\,\,\cos\eta_2 & \sin\eta_2 & 0\\
-\sin\eta_2 & \cos\eta_2 & 0\\
\qquad 0 & 0 & e^{-i\kappa_2}\end{array}\right),\quad
b=-\left(\begin{array}{lcr}\,\,\,\,\,\cos\eta_1 & \sin\eta_1 & 0\\
-\sin\eta_1 & \cos\eta_1 & 0\\
\qquad 0 & 0 & e^{-i\kappa_1}\end{array}\right),
\label{eq_matricesab}
\end{equation}
where $\kappa_1$ and $\kappa_2$ are mixing phases between active and sterile
neutrinos, whereas $\eta_1$ and $\eta_2$ are mixing angles between them. The matrix
$a$ in the form of equation~(\ref{eq_matricesab}) was proposed in
Ref.~\citen{KhruFom2016}. In order to make our calculations more specific, we
will use the following sample values for new mixing parameters:
\begin{equation}
\kappa_1=\kappa_2=-\pi/2,\quad \eta_1=5^{\circ},\quad \eta_2=\pm 30^{\circ},
\label{eq_etakappa}
\end{equation}
and assume that the small parameter $\epsilon$ satisfies at least the
condition $\epsilon\lesssim 0.03$.

The neutrino masses will be given by a normally ordered set of values
$\{m\}=\{m_i,m_{i'}\}$. For active neutrinos we will use the neutrino mass
estimations, which were proposed in
Refs.~\citen{Zysina2014,KhruFom2016,PAZH2016} for NO-case (in units of eV) and
which do not contradict to the known experimental data up to now.
\begin{equation}
m_1\approx 0.0016, \quad m_2\approx 0.0088, \quad m_3\approx 0.0497\,.
\label{eq_activmasses}
\end{equation}
The values of the mixing angles $\theta_{ij}$ of active neutrinos that
determine the Pontecorvo--Maki--Nakagawa--Sakata mixing matrix will be taken
from relations $\sin^2\theta_{12}\approx 0.318$,
$\sin^2\theta_{23}\approx 0.566$ and $\sin^2\theta_{13}\approx 0.0222$,
which are obtained from the processing of experimental data for NO and given in
Ref.~\citen{salas}.

In Ref.~\citen{khfo} the version of the Light Mass Option (LMO1 version)  of the (3+1+2) model has been considered for $m_4$, $m_5$, and $m_6$ mass values:
\begin{equation}
\{m\}_{\rm LMO1}=\{1.1,\,1.5\!\times\!10^3,\,7.5\!\times\!10^3 \}.
\label{eq_LMO1}
\end{equation}
In order to reproduce in more detail the electrons energy spectrum observed in the XENON1T experiment in what follows we choose a comparatively higher  mass $m_5$, 
than the corresponding mass value given in Ref.~\citen{khfo} (see (\ref{eq_LMO1})).
The  $m_4$ and practically $m_6$ mass values are unchanged, furthermore the  $m_4$ value meets currently available constraints ~\cite{archi,vag}. Thus, below we will use the following  $m_4$, $m_5$, and $m_6$ mass values for sterile mass states:
\begin{equation}
\{m\}_{\rm LMO}=\{1.1,\,3.4\!\times\!10^3,\,7.6\!\times\!10^3 \}.
\label{eq_LMO}
\end{equation}

With the LMO set of the mass values above it remains possible to explain the appearance of anomalies at short distances in neutrino data \cite{Gariazzo2017}.
Note that sterile
neutrinos with masses about several keVs are also used for interpretation
of some astrophysical data \cite{asch18}, so this adds considerable support for
our choice of the $m_5$ mass value as $3.4$~keV and the $m_6$ mass value as $7.6$~keV. 

\section{Data relevant to the electronic recoil events excess in the XENON1T experiment and their interpretation in the context of the (3+1+2) model}
\label{xenon_ex}

Recently the XENON1T experiment data have been reported on the observation of the excess of electronic recoil events in the energy region between 1 and 7 keV \cite{apr}. 
The XENON1T experiment operated underground at the INFN Laboratori Nazionali del Gran Sasso.  This experiment, employing a liquid-xenon time projection chamber with a
2.0-tonne active target, was primarily designed to detect Weakly Interacting Massive
Particle (WIMP) dark matter. A particle interaction within the detector produces both
prompt scintillation and delayed electroluminesence signals. These light signals are detected by arrays of photomultiplier tubes on the top and bottom of the active volume, and are used to determine the deposited energy and interaction position of an event. The ratio between delayed electroluminesence signals and prompt scintillation signals is used to distinguish electronic recoils, produced by, e.g., gamma rays or beta electrons, from nuclear recoils, produced by, e.g., neutrons or WIMPs, allowing for a degree of particle identification.

In what follows we focus on the possibility of describing, in the
framework of the (3+1+2) model considered above, the excess of electronic recoil events observed 
in the  XENON1T experiment . We suggest that this excess can be naturally attributed
 to interaction of electrons with dark bosons arose for the most part in decay processes of hidden and dark neutrinos. Note that these processes can only slightly produce  photons as well.
 A plausible mechanism for photon appearance can be a kinetic mixing to only a small extent       between a photon and a dark boson \cite{hold}. It is assumed that hidden and dark neutrinos originally possess of nonrelativistic velocities and the dark boson have a very small mass.   So dark bosons and photons  can be emitted in transitions among  mass component parts of
 dark, hidden and sterile neutrinos assuming that the  sterile neutrino,
 which is mainly the $m_4$ mass state, is practically stable. Thus using this approach we predict three peaks in the 1 -- 7 keV energy region of electronic recoil events at energies about $1.7$ keV, $3$ keV and $3.8$ keV. This prediction can be tested as in the XENON1T experiment when a high-statistics data set will be available as in future  experiments of this kind. 
Note that the used above the LMO variant of the (3+1+2) neutrino model with the decaying heavy  neutrinos and the light stable sterile neutrino still remain operable for description of the SBL neutrino anomalies (see, e.g., \cite{khfo,mona,dego,abdu}).

\section{Discussion and conclusions}
\label{Section_Conclusion}

In this paper, we use the phenomenological (3+1+2) neutrino model with
three active and three sterile neutrinos for description of the excess of electronic
recoil events in the 1 -- 7 keV energy region  found in the data of the XENON1T experiment \cite{apr}. This excess can be naturally attributed  to interaction of electrons with dark bosons and photons emitted in decays of the sterile neutrino mass states with the masses
 $m_5=3.4$ keV and $m_6=7.6$ keV while the sterile neutrino mass state with the mass
 $m_4=1.1$ eV is practically stable. In the context of this approach three peaks in the 1 -- 7 keV energy region of electronic
recoil events at energies  are predicted. These predictions will
be tested as in the XENON1T experiment as in future experiments, such as the upcoming PandaX-4T \cite{panda}, LZ \cite{lz} and XENONnT \cite{apr} experiments.

 The possible existence of the three  massive  sterile neutrinos may have a perceptible  influence on some phenomena in neutrino physics, astrophysics and cosmology. By way of illustration we refer to the possibility to interpret the SBL anomalies data in the framework of the (3+1+2)  model with sterile neutrinos \cite{khfo}. Moreover the incorporation of two decaying sterile neutrinos with $3.4$~keV and  $7.6$~keV masses allows us 
% to explain the registration of the line $3.55$~keV in the gamma spectra of some %astrophysical sources \cite{Bulb2014,Boya2014,Horiuchi}, as well as 
to predict amplification or appearance of the lines in the range of several keVs in the gamma spectra of some astrophysical sources. The presence of stable sterile neutrino mass state with the mass about $1$~eV will make an impact on a value of the important cosmological parameter $ \Delta N_{eff}$, 
besides it is possible to some extent this can matter for the resolution of the issue 
concerning the $ H_{0}$ tension \cite{archi,vag}.

\end{document}